\documentclass[12pt]{article}

\usepackage{amsmath,amsfonts,amssymb}

\usepackage{subfigure}

\usepackage{epsfig,graphicx,bm}

\usepackage{epstopdf}

\usepackage{lipsum}

\begin{document}
	
	\begin{titlepage}
		
		\begin{center}
			
			\vskip 0.4 cm

				{\Large \bf Action for $N$ D0-Branes Invariant Under Gauged Galilean
				Transformations}

			\vskip 1cm
			
			\vspace{1em}J. Kluso\v{n},		
			\footnote{Email addresses:	 klu@physics.muni.cz   (J.Kluso\v{n}), }\\
			
			\vspace{1em}
\textit{Department of Theoretical Physics and
				Astrophysics, Faculty of Science,\\
				Masaryk University, Kotl\'a\v{r}sk\'a 2, 611 37, Brno, Czech
				Republic}
			
			\vskip 0.8cm
			
		\end{center}
		
		\begin{abstract}
In this short note we formulate an action for $N$ D0-branes that is manifestly invariant under gauged Galilean transformations. We also find its canonical form and determine first class constraints that are generators of gauge transformations.

		\end{abstract}
	\end{titlepage}
	
	\bigskip
	
	\newpage

\def\bn{\mathbf{n}}
\newcommand{\bC}{\mathbf{C}}
\newcommand{\bD}{\mathbf{D}}
\def\hf{\hat{f}}
\def\tK{\tilde{K}}
\def\mC{\mathcal{C}}
\def\bJ{\mathbf{J}}
\def\bk{\mathbf{k}}
\def\tr{\mathrm{tr}\, }
\def\tmH{\tilde{\mH}}
\def\tY{\mathcal{Y}}
\def\nn{\nonumber \\}
\def\bI{\mathbf{I}}
\def\tmV{\tilde{\mV}}
\def\e{\mathrm{e}}
\def\bP{\mathbf{P}}
\def\bE{\mathbf{E}}
\def\bX{\mathbf{X}}
\def\bY{\mathbf{Y}}
\def\bR{\bar{R}}
\def\hN{\hat{N}}
\def\hK{\hat{K}}
\def\hnabla{\hat{\nabla}}
\def\hc{\hat{c}}
\def\mH{\mathcal{H}}
\def \Gi{\left(G^{-1}\right)}
\def\hZ{\hat{Z}}
\def\bz{\mathbf{z}}
\def\bK{\mathbf{K}}
\def\iD{\left(D^{-1}\right)}
\def\tmJ{\tilde{\mathcal{J}}}
\def\tr{\mathrm{Tr}}
\def\mJ{\mathcal{J}}
\def\partt{\partial_t}
\def\parts{\partial_\sigma}
\def\bG{\mathbf{G}}
\def\str{\mathrm{Str}}
\def\Pf{\mathrm{Pf}}
\def\bM{\mathbf{M}}
\def\tA{\tilde{A}}
\newcommand{\mW}{\mathcal{W}}
\def\bx{\mathbf{x}}
\def\by{\mathbf{y}}
\def \mD{\mathcal{D}}
\newcommand{\tZ}{\tilde{Z}}
\newcommand{\tW}{\tilde{W}}
\newcommand{\tmD}{\tilde{\mathcal{D}}}
\newcommand{\tN}{\tilde{N}}
\newcommand{\hC}{\hat{C}}
\newcommand{\hg}{g}
\newcommand{\hX}{\hat{X}}
\newcommand{\bQ}{\mathbf{Q}}
\newcommand{\hd}{\hat{d}}
\newcommand{\tX}{\tilde{X}}
\newcommand{\calg}{\mathcal{G}}
\newcommand{\calgi}{\left(\calg^{-1}\right)}
\newcommand{\hsigma}{\hat{\sigma}}
\newcommand{\hx}{\hat{x}}
\newcommand{\tchi}{\tilde{\chi}}
\newcommand{\mA}{\mathcal{A}}
\newcommand{\ha}{\hat{a}}
\newcommand{\tB}{\tilde{B}}
\newcommand{\hrho}{\hat{\rho}}
\newcommand{\hh}{\hat{h}}
\newcommand{\homega}{\hat{\omega}}
\newcommand{\mK}{\mathcal{K}}
\newcommand{\hmK}{\hat{\mK}}
\newcommand{\hA}{\hat{A}}
\newcommand{\mF}{\mathcal{F}}
\newcommand{\hmF}{\hat{\mF}}
\newcommand{\tk}{\tilde{k}}
\newcommand{\hQ}{\hat{Q}}
\newcommand{\mU}{\mathcal{U}}
\newcommand{\hPhi}{\hat{\Phi}}
\newcommand{\hPi}{\hat{\Pi}}
\newcommand{\hD}{\hat{D}}
\newcommand{\hb}{\hat{b}}
\def\I{\mathbf{I}}
\def\tW{\tilde{W}}
\newcommand{\tD}{\tilde{D}}
\newcommand{\mG}{\mathcal{G}}
\def\IT{\I_{\Phi,\Phi',T}}
\def \cit{\IT^{\dag}}
\newcommand{\hk}{\hat{k}}
\def \cdt{\overline{\tilde{D}T}}
\def \dt{\tilde{D}T}
\def\bra #1{\left<#1\right|}
\def\ket #1{\left|#1\right>}
\def\mV{\mathcal{V}}
\def\Xn #1{X^{(#1)}}
\newcommand{\Xni}[2] {X^{(#1)#2}}
\newcommand{\bAn}[1] {\mathbf{A}^{(#1)}}
\def \bAi{\left(\mathbf{A}^{-1}\right)}
\newcommand{\bAni}[1]
{\left(\mathbf{A}_{(#1)}^{-1}\right)}
\def \bA{\mathbf{A}}
\newcommand{\bT}{\mathbf{T}}
\def\bmR{\bar{\mR}}
\newcommand{\mL}{\mathcal{L}}
\newcommand{\mbQ}{\mathbf{Q}}
\def\mat{\tilde{\mathbf{a}}}
\def\mtF{\tilde{\mathcal{F}}}
\def \tZ{\tilde{Z}}
\def\mtC{\tilde{C}}
\def \tY{\tilde{Y}}
\def\pb #1{\left\{#1\right\}}
\newcommand{\E}[3]{E_{(#1)#2}^{ \quad #3}}
\newcommand{\p}[1]{p_{(#1)}}
\newcommand{\hEn}[3]{\hat{E}_{(#1)#2}^{ \quad #3}}
\def\mbPhi{\mathbf{\Phi}}
\def\tg{\tilde{g}}
\newcommand{\phys}{\mathrm{phys}}

\section{Introduction and Summary}
Relational mechanics is formulation of dynamics of particles that is closely 
related to Mach's idea that claims that dynamics of $N$ particles should be a theory 
of relations about these quantities without any reference to external non-material entities. The question is how to make these ideas more concrete. One such a possibility
was proposal that the Lagrangian should be invariant under gauged Galilean group 
\cite{Barbour:1982gha}
\footnote{See also \cite{Barbour:2010dp,Lynden-Bell:1995cmj}.}. Such Lagrangian can be found when  its measure (kinetic  energy term) is replaced  
 with a measure that is defined in the space of orbits, where orbits correspond to a set of configurations which are  equivalent under gauge transformations. Then it was shown that the
solutions of this gauge invariant dynamics correspond to the solutions of the original Lagrangian
with vanishing total momentum and angular momentum. An alternative proposal how to construct relational mechanics was presented in \cite{Ferraro:2014yza}. This construction starts with the original Lagrangian invariant under rigid Galilean transformation. As the next step we  add to it specific counterterms that compensate changes in the kinetic energy term under time dependent Galilean transformations. As a result Lagrangian invariant under time dependent Galilean transformation was derived in
\cite{Ferraro:2014yza} and it was also shown that corresponding equations of motion are valid in any frame making concrete implementation of the Mach's principle. 

In more details, Newton's mechanics is invariant under time independent translations, 
space translations and rotations of particle's positions $\bx_i$
\begin{eqnarray}\label{tr1}
&&	t'=t+\epsilon \ , \quad  \epsilon=\mathrm{const} \ , \nonumber \\
&&	\bx'_i=\bx_i+\xi \ , \quad \xi=\mathrm{const} \ , \nonumber \\
&&	\bx'_i=\bA \bx_i \ , \nonumber \\
\end{eqnarray}
where $\bA$ is orthogonal matrix, $i=1,\dots,N$ where $i$ labels 
particles in ansamble. Further, Newton's laws are invariant under Galilean transformations
\begin{equation}\label{tr2}
\bx'_i=\bx_i+\mathbf{V} t \ , 
\end{equation}
which are special case of local time dependent translations when we identify $\xi=\mathbf{V}t$.
The transformations (\ref{tr1}) and (\ref{tr2}) represent Galilean group of Newtonian mechanics. 
In fact, there is a privileged set of inertial frames and clocks in Newtonian mechanics which are related each other through the transformation of the Galilean group. Then we leave an idea of these privileged frames when the Galilean group of transformation becomes gauge group with time dependent parameters. We can also eliminate the absolute time when we gauge time translation as $t'=t+\epsilon(t)$. In such a formulation of mechanics there are no privileged frames and clocks and 
it becomes purely relational. 

Detailed analysis of formulation of relational mechanics was performed in \cite{Ferraro:2014yza} where systems of $N$ non-relativistic particles was studied. The Lagrangian that is invariant under time dependent Galilean group was found there together with corresponding Hamiltonian formulation. It was also shown there that Relational Mechanics contain frames where Newtonian mechanics is valid and these frames are determined by mass distribution of all particles which is an essence of the Mach's principle.

In this article we would like to follow the procedure used in  \cite{Ferraro:2014yza} and in 
\cite{Glampedakis:2022fqu}  in case of one very interesting non-relativistic system which is low energy Lagrangian for $N$ D0-branes in string theory 
\cite{Polchinski:1995mt,Dai:1989ua}. It is well known that 
the low-energy Lagrangian for a system of many type IIA D0-branes is the matrix
quantum mechanics Lagrangian arising from the dimensional reduction to $0 + 1$ dimensions
of the $10D$ super Yang-Mills Lagrangian, for review see 
for example  \cite{Sen:1998kr,Taylor:1997dy,Taylor:1999qk}. An importance of this action is that it is the key point in the formulation of Matrix theory \cite{Banks:1996vh} which is strictly speaking defined for infinite number of D0-branes even if there is a version of Matrix theory that is correct for finite $N$ as well \cite{Seiberg:1997ad,Sen:1997we}. Now we show that it is possible to formulate Lagrangian for $N$ D0-branes that is manifestly invariant under gauged Galilean transformations and hence, according to the extended discussion presented in 
\cite{Ferraro:2014yza} corresponds to relational mechanics for $N$ D0-branes. On the other hand  it is not possible to write it in manifestly relational form where the Lagrangian depends on relative 
velocities and distances of particles in the general case due to the  fact that fundamental objects in matrix theory 
are matrices rather than coordinates of individual  D0-branes. On the other hand this can be easily done in the approximation when off-diagonal components of matrices are small with respect to the diagonal ones so that we can neglect them and we show that in this case the Lagrangian takes purely relational form.

As the next step in our research  we find canonical form of this theory and we identify two sets of the generators of the first class constraints. We also  show that by appropriate fixing the gauge symmetry this theory reduces to the ordinary finite Matrix theory. 

 We mean that this is nice and interesting result that should be developed further. 
In fact, the natural and very important question is to include fermionic terms in the constructions of relational mechanics for $N$ D0-branes. The second question is whether it is possible to  generalize this  construction of relational mechanics  to the case of 
the full non-linear version of the action as it is represented by non-abelian Dirac-Born-Infeld action for $N$ D0-branes \cite{Tseytlin:1997csa,Myers:1999ps}. We hope to return to these problems in near future. 

The structure of this paper is as follows. In the next section (\ref{second}) we find an action for $N$ D0-branes that is invariant under gauged Galilean transformations. 
Then in section (\ref{third}) we determine its canonical form and identify constraints structure of the theory. Finally in section (\ref{fourth}) we show how it is possible to make this theory manifestly reparametrization invariant.

 \section{Relational Formulation of Action for $N$ D0-Branes}
 \label{second}
 In this section we find the form of low energy effective action for $N$ D0-branes
 that is invariant under gauged  Galilean transformations. We start with the Lagrangian for $N$ D0-branes that at the leading order corresponds to 
 $U(N)$ Super-Yang-Mills mechanical system that has the form
\begin{equation}\label{LagD0}
S=\int dt L \ , \quad 
L=\frac{1}{2gl_s}\tr\left[\dot{\Phi}^I
\dot{\Phi}^I+\frac{1}{2}[\Phi^I,\Phi^J][\Phi^I,\Phi^J]+\mathrm{fermions}\right] \ , 
\end{equation}
where $\Phi^I_{ij}$ are $N\times N$ Hermitean matrices where $I,J=1,2,\dots,9$ and where $i,j,\dots=1,\dots,N$. Further, $g_s$ is string coupling constant and $l_s$ is the string length. Note that the transverse space is nine-dimensional Euclidean space and repeated indices mean summation over them. Finally this Lagrangian contains terms with fermions that makes the Lagrangian $N=16$ Super-Yang-Mills mechanics. In more details, the Lagrangian (\ref{LagD0}) can be defined 
by dimensional reduction of $10$ dimensional $\mathcal{N}=1$ super-Yang-Mills theory 
with gauge group $U(N)$ to the $0+1$ dimensions.
 In what follows we restrict ourselves
to the bosonic terms only leaving its extension to the fully supersymmetric invariant case in the near future. 

 The action (\ref{LagD0}) is invariant under rigid  translation
\begin{equation}
\Phi'^I(t)=\Phi^I(t)+\xi^I \bI_{N\times N} \ , 
\end{equation}
where $\xi^I$ is constant and where $\bI_{N\times N}$ is unit $N\times N$ matrix. Further, the Lagrangian is invariant under rigid 
rotation 
\begin{equation}\label{rigrot}
\Phi^I(t)=\Lambda^I_{ \ J}\Phi^J(t) \ , 
\end{equation}
where $\Lambda^I_{ \ J}$ obey the relations
\begin{equation}\label{defLambda}
\Lambda^I_{ \ K}\delta_{IJ}\Lambda^J_{ \ L}=\delta_{KL} \ . 
\end{equation}
Let us now construct Lagrangian that is invariant under time dependent translation 
\begin{equation}\label{transtr}
\Phi'^I(t)=\Phi^I(t)+\xi^I(t) \bI_{N\times N} \ .
\end{equation}
Clearly the potential term is invariant under this transformation while the kinetic 
term transforms as
\begin{eqnarray}
\delta \left(\frac{1}{2g_sl_s}\tr [\dot{\Phi}^I\dot{\Phi}^I]\right)
=\frac{1}{g_sl_s}\tr \delta \dot{\Phi}^I\dot{\Phi}^I=
\frac{1}{g_s l_s}\dot{\xi}^I\tr \dot{\Phi}^I \ . 
\nonumber \\
\end{eqnarray}
In order to compensate this transformation we consider following variation
\begin{eqnarray}
\frac{1}{2Ng_sl_s}\delta (\tr\dot{\Phi}^I\tr\dot{\Phi}^I)
=\frac{1}{gl_s}\dot{\xi}^I\tr \dot{\Phi}^I
\nonumber \\
\end{eqnarray}
so that following combination 
\begin{equation}\label{comtr}
\frac{1}{2g_sl_s}\tr [\dot{\Phi}^I\dot{\Phi}_I]-
\frac{1}{2Ng_s l_s }\tr \dot{\Phi}^I\tr \dot{\Phi}^I
\end{equation}
is invariant under the gauge symmetry (\ref{transtr}).
As the next step we proceed to the analysis of the invariance of the Lagrangian 
under time dependent rotation.
\\
{\bf Time Dependent Rotation}

As we have argued above the action (\ref{LagD0}) is invariant also under rigid rotation 
(\ref{rigrot}) where $\Lambda^I_{ \ J}$ obey the relation 
(\ref{defLambda}). It is convenient to write it in infinitesimal form when 
we define 
\begin{equation}\label{lambdasmall}
\Lambda^I_{ \ J}=\delta^I_{ \ I}+\omega^I_{ \ J}  \ , \quad \omega^I_{ \ J}\ll \delta^I_J \ . 
\end{equation}
Then (\ref{defLambda}) implies 
\begin{equation}
\delta_{KJ}\omega^J_{ \ L}+\delta_{LI}\omega^I_{ \ K}=0 
\Rightarrow \omega_{KL}+\omega_{LK}=0 \ .
\end{equation}
Let us now presume that $\omega^I_{ \ J}$ depend on $t$ so that 
 (\ref{lambdasmall}) 
 gives
 \begin{equation}
\delta \Phi^I=\Phi'^I-\Phi^I=\omega^I_{ \ J}\Phi^J
\ , \quad 
\delta \dot{\Phi}^I\equiv \dot{\omega}^I_{ \ J}\Phi^J+
\omega^I_{\ J}\dot{\Phi}^J \ .
\end{equation}
Then  the  variation of the kinetic term is equal to
\begin{eqnarray}
&&\frac{1}{2g_sl_s}\delta \tr(\dot{\Phi}^I\delta_{IJ}\dot{\Phi}^J)
=\frac{1}{g_sl_s}\dot{\omega}^I_{ \ K}
\tr (\Phi^K\delta_{IJ}\dot{\Phi}^J)+
\frac{1}{g_sl_s}\tr( \dot{\Phi}^J\omega_{JK}\dot{\Phi}^K)
\nonumber \\
&&=\frac{1}{g_sl_s}\dot{\omega}^I_{ \ K}
\tr (\Phi^K\delta_{IJ}\dot{\Phi}^J) \ ,  \nonumber \\
\end{eqnarray}
where the last term on the first line vanishes due to the anti-symmetry of $\omega_{IJ}$.
As the next step we should calculate the variation of  new compensating term and we find 
\begin{eqnarray}
&&-\delta \left(\frac{1}{2Ng_s l_s}\tr \dot{\Phi}^I\tr \dot{\Phi}^I\right)
=-\dot{\omega}^I_{ \ K}\frac{1}{Ng_sl_s}\tr \Phi^K
\delta_{IJ}\tr \dot{\Phi}^J-
\tr\frac{1}{Ng_sl_s}\omega_{KJ}\tr \dot{\Phi}^K\tr \dot{\Phi}^J=\nonumber \\ 
&&=-\dot{\omega}^I_{ \ K}\frac{1}{Ng_sl_s}\tr \Phi^K 
\delta_{IJ}\tr \dot{\Phi}^J \ ,  \nonumber \\
\end{eqnarray}
where again the last term on the first line vanishes due to the anti-symmetry of $\omega_{IJ}$.
In other words the variation of the kinetic term is equal to
\begin{eqnarray}
&&\delta\left(\frac{1}{2g_sl_s} \tr(\dot{\Phi}^I\delta_{IJ}\dot{\Phi}^J)
-\frac{1}{2Ng_s l_s}\tr \dot{\Phi}^I\tr \dot{\Phi}^I\right)=
\dot{\omega}_{JK}\bJ^{KJ} \ , \nonumber \\
&&  \bJ^{IJ}=\frac{1}{2g_sl_s}\tr (\Phi^I\dot{\Phi}^J-\Phi^J\dot{\Phi}^I)-
\frac{1}{2g_s l_sN}(\tr \Phi^I
\tr \dot{\Phi}^J-\tr\Phi^J\tr\dot{\Phi}^I)=-\bJ^{JI} \ . \nonumber \\
 \end{eqnarray}
Note that $\bJ^{IJ}$ transforms under time dependent rotation as 
\begin{eqnarray}
\delta \bJ^{IJ}=
\omega^I_{ \ K}\bJ^{KJ}+\bJ^{IK}\omega_K^{ \ J}
-\dot{\omega}^I_{ \ K}I^{KJ}-I^{IK}\dot{\omega}_K^{ \ J} \ ,  \nonumber \\
\end{eqnarray}
where 
\begin{equation}
I^{KJ}=\frac{1}{2g_sl_s}\tr (\Phi^K\Phi^J)-\frac{1}{2g_sl_s N}\tr \Phi^K
\tr\Phi^J \ .
\end{equation}
Our goal is to add new additional term to the Lagrangian to make it invariant
under time dependent rotation. We 
propose that such term has a form
\begin{equation}\label{newterm}
-\frac{1}{2}\bJ^{IJ}\bM_{IJ,KL}\bJ^{KL} \ , 
\end{equation}
where the matrix $\bM_{IJ,KL}$ is matrix in $I,J,K,L$ indices while it is scalar
with respect to the $U(N)$ structure. 

Now we proceed to the construction of the object $\bM_{IJ,KL}$. First of all we demand that it does not depend on time derivative of $\Phi$ so that it transforms as ordinary tensor  under time dependent rotation. Then under time dependent rotation the new term (\ref{newterm})  transforms as
\begin{eqnarray}
&&\delta(\frac{1}{2}\bJ^{IJ}\bM_{IJ,KL}\bJ^{KL})=
-\dot{\omega}^I_{ \ M}I^{MJ}\bM_{IJ,KL}\bJ^{KL}
-I^{IM}\dot{\omega}_M^{ \ J}\bM_{IJ,KL}\bJ^{KL}=
\nonumber \\ 
&&
=-\dot{\omega}^I_{ \ M}I^{MJ}(M_{IJ,KL}-M_{JI,KL})\bJ^{KL}=2\dot{\omega}^I_{ \ M}
I^{MJ}\bM_{JI,KL}\bJ^{KL} \ , 
\nonumber \\
\end{eqnarray}
where we presume that $\bM_{IJ,KL}=-\bM_{JI,KL}$ as a consequence of  the fact that
$\bJ^{IJ}=-\bJ^{JI}$. In order to cancel variation of the kinetic term we should require that 
\begin{equation}\label{var}
2I^{MJ}\bM_{JI,KL}=\delta_{IL} \delta_K^M \ . 
\end{equation}
To proceed further we introduce $I^{-1}_{IJ}$ as matrix inverse to 
$I^{JK}$ so that $I^{IJ}I^{-1}_{JK}=\delta^I_K$. Then if we multiply 
(\ref{var})
with $(I^{-1})_{RM}$ we get
\begin{equation}
\bM_{RI,KL}=\frac{1}{2}\delta_{IL} I^{-1}_{RK} \ . 
\end{equation}
We see that  it is natural to define matrix  $\bM_{IJ,KL}$ as
\begin{eqnarray}
	\bM_{IJ,KL}=\frac{1}{2}\delta_{JL}I^{-1}_{IK} \ , \quad 
	\bM_{JI,KL}=-\frac{1}{2}\delta_{JL}I^{-1}_{IK} \ , \nonumber \\
	\bM_{KL,IJ}=\frac{1}{2}\delta_{LJ}I^{-1}_{KI} \ , \quad 
	\bM_{IJ,LK}=-\frac{1}{2}\delta_{JL}I^{-1}_{IK} \ . \nonumber \\
	\nonumber \\
\end{eqnarray}

In summary, we have found Lagrangian for $N$ D0-branes that is invariant under local translation and rotation and that has the form 
\begin{equation}\label{LD0}
	L=\frac{1}{2gl_s}\tr\left[\dot{\Phi}^I
	\dot{\Phi}_I+\frac{1}{2}[\Phi^I,\Phi^J][\Phi^I,\Phi^J]
	\right]-\frac{1}{2Ng_s l_s}\tr \dot{\Phi}^I\dot{\Phi}^I-
	\frac{1}{2}\bJ^{IJ}\bM_{IJ,KL}\bJ^{KL} \ . 
\end{equation}
This is final form of the Lagrangian for $N$ D0-branes that is invariant under time dependent Galilean transformation so that
this Lagrangian is valid in any frame. In fact, following \cite{Ferraro:2014yza}
we can interpret is as relational formulation of D0-brane mechanics. To see this more clearly 
let us consider
situation when the matrices $\Phi^I$ are diagonal, or say alternatively, situation when  we can neglect all 
off diagonal terms with respect to diagonal ones. Then the matrices $\Phi^I$ have the form
\begin{equation}
	\Phi^I_{ij}=x^I_i\delta_{ij} \ , 
\end{equation}
where $x^I_i$ are coordinates of individual $i-$th D0-brane. 
With such a configuration we find that the potential term vanishes while the kinetic term has the form
\begin{equation}
	\frac{1}{2g_s l_s}
	(\sum_i v_i^Iv_i^I-\frac{1}{N}\sum_i v_i^I\sum_j v_j^I) \ , \quad v^I_i=\frac{dx^I_i}{dt}
\end{equation}
that can be written in an alternative form 
\begin{equation}
	\frac{1}{4Ng_s l_s}
	\sum_{i,j}(v_i^I-v_j^I)(v_i^I-v_j^I)
\end{equation}
which nicely demonstrate the relational form of this Lagrangian. Further, matrix $I^{IJ}$ has the form
\begin{equation}\label{Irel}
I^{IJ}=\frac{1}{2g_sl_s}(\sum_i x^I_i x^J_i-\sum_i x^I_i\sum_j x^J_j)=
\frac{1}{4g_s l_s N}\sum_{i,j}(x_i^I-x_j^I)(x^J_i-x^J_j) \ . 
\end{equation}
In the same way we proceed with $\bJ^{IJ}$ and we get
\begin{eqnarray}\label{bJrel}
&&\bJ^{IJ}=\frac{1}{2g_sl_s}\sum_i(x^I_iv^J_i-x^J_iv^I_i)
-\frac{1}{2g_s l_sN}(\sum_i x^I_i\sum_j v^J-\sum_i x^J_i\sum_j v^I_j)=
\nonumber \\
&&=\frac{1}{4g_sl_sN}
\sum_i\sum_j\left((x^I_i-x^I_j)(v^J_i-v^J_j)-
(x^J_i-x^J_j)(v^I_i-v^I_j)\right) \  \nonumber \\
\end{eqnarray}
which again depends on relative distances and velocities.
  In summary we obtain Lagrangian
\begin{eqnarray}
L=	\frac{1}{4Ng_s l_s}
\sum_{i,j}(v_i^I-v_j^I)(v_i^I-v_j^I)-\frac{1}{2}\bJ^{IJ}\bM_{IJ,KL}
\bJ^{KL} \ 
\nonumber \\
\end{eqnarray}
that has manifestly form of relational mechanics as follows from 
 (\ref{Irel}) and (\ref{bJrel}).

\section{Hamiltonian Formalism}\label{third}
In this section we find Hamiltonian from Lagrangian (\ref{LD0}). 
In the first step we introduce conjugate momenta to the matrix elements $\Phi^I_{ij}$ where we will tread matrix elements as independent keeping in mind that we have $\Phi_{ij}=\Phi_{ji}^*$. Then from (\ref{LD0}) we obtain
\begin{eqnarray}\label{Piij}
&&	(\Pi_I)_{ij}=\frac{\delta L}{\delta \dot{\Phi}^I_{ij}}=
	\frac{1}{g_s l_s}(\dot{\Phi}_I)_{ji}-\frac{1}{Ng_s l_s}\delta_{ji}\tr{\dot{\Phi}_I}-\nonumber \\
&&\frac{1}{2g_sl_s}(\Phi^K_{ji}\delta^L_I-\Phi^L_{ji}\delta^K_I)-
\frac{1}{N}(\tr \Phi^K\delta_{ji}\delta^L_I-\tr \Phi^L\delta_{ji}\delta^K_I)	
)	\bM_{KL,MN}\bJ^{MN} \ , \nonumber \\
\end{eqnarray}
using
\begin{equation}
	\frac{\delta \bJ^{KL}}{\delta \dot{\Phi}^I_{ij}}=
\frac{1}{2g_sl_s}(\Phi^K_{ji}\delta^L_I-\Phi^L_{ji}\delta^K_I)-
\frac{1}{2g_sl_s N}(\tr \Phi^K\delta_{ji}\delta^L_I-\tr \Phi^L\delta_{ji}\delta^K_I)	\ . 
\end{equation}
As the next step we define Hamiltonian in the standard way
\begin{eqnarray}
&&H=(\Pi_I)_{ij}\dot{\Phi}^I_{ij}-L \nonumber \\
&&=\frac{1}{2g_sl_s}\dot{\Phi}^I_{ij}
\dot{\Phi}^I_{ji}-\frac{1}{2g_sl_sN}\tr \dot{\Phi}^I
\tr\dot{\Phi}^I-\frac{1}{2}\bJ^{IJ}\bM_{IJ,KL}\bJ^{KL}-\frac{1}{4g_sl_s}\tr[\Phi^I,\Phi^J]
[\Phi^I,\Phi^J] \ . 
\nonumber \\
\end{eqnarray}
To proceed further note  that (\ref{Piij}) implies 
\begin{eqnarray}
\bP_I\equiv\tr \Pi_I=(\Pi_I)_{ij}\delta_{ji}=0 \ , 
\nonumber \\
\end{eqnarray}
so that $\bP_I\approx 0$ is primary constraint of the theory. Further, from  (\ref{Piij}) we also get
\begin{eqnarray}
	\Phi^I_{ij}(\Pi_J)_{ij}-\Phi^J_{ij}(\Pi_I)_{ij}=0
	\nonumber \\
\end{eqnarray}
so that there is second set of  primary constraints 
\begin{equation}
	\bJ^{IJ}= \Phi^I_{ij}\Pi^J_{ij}- \Phi^J_{ij}\Pi^I_{ij}\approx 0 \ , 
\end{equation}
where again repeated indices mean summation over them. 

Now we should check that they are the first class constraints. To do this we introduce
canonical Poisson brackets
\begin{equation}
	\pb{\Phi^I_{ij},(\Pi_J)_{kl}}=\delta^I_J\delta_{ik}\delta_{jl} \ . 
\end{equation}
Clearly we have
\begin{eqnarray}
	\pb{\bP_I,\bP_J}=0
\end{eqnarray}
and also
\begin{eqnarray}
	\pb{\bJ^{IJ},\bP_K}=\delta^I_K\tr \Pi^J-\delta^J_K\tr \Pi^I
=\delta^I_K\bP^J-\delta^J_K\bP^I\approx 0 \ .\nonumber \\
\end{eqnarray}
As the last Poisson bracket we calculate $\pb{\bJ^{IJ},\bJ^{KL}}$ and we obtain
\begin{eqnarray}
&&\pb{\bJ^{IJ},\bJ^{KL}}=\delta^{IL}\Pi^J_{ij}\Phi^K_{ij}-\delta^{JK}\Phi^I_{mn}\Pi^L_{mn}-\delta^{JL}\Pi^I_{mn}\Phi^K_{mn}+\delta^{IK}\Phi_{mn}^J\Pi^L_{mn}-\nonumber \\
&&-\delta^{IK}\Phi_{mn}^L\Pi_{mn}^J+\delta^{JL}\Phi^I_{mn}\Pi^K_{mn}+
\delta^{JK}\Pi^I_{mn}\Phi^L_{mn}-\delta^{IL}\Phi^J_{mn}\Pi^K_{mn}=
\nonumber \\
&&=\delta^{IK}\bJ^{JL}-
\delta^{IL}\bJ^{JK}-\delta^{JK}\bJ^{IL}+\delta^{JL}\bJ^{IK} \approx 0 \ . \nonumber \\
\end{eqnarray}
In summary we find that $\bP_I\approx 0 \ , \bJ^{IJ}\approx 0$ are first class constraints. 
We will discuss their properties below. 

Finally we return to the Hamiltonian and express it in the form of canonical variables.
Using (\ref{Piij}) we obtain
\begin{eqnarray}
(\Pi_I)_{ij}(\Pi_I)_{ji}
=\frac{1}{g_s^2l_s^2}\tr \dot{\Phi}^I\dot{\Phi}_I-\frac{1}{N g_s^2l_s^2}\tr\dot{\Phi}_I
\tr\dot{\Phi}_I-\frac{2}{g_sl_s}\bJ^{KL}\bM_{KL,MN}\bJ^{MN}
+\frac{1}{g_sl_s}\bJ^{KL}\bM_{KL,MN}\bJ^{MN}\nonumber \\
\end{eqnarray}
and  we find that the bare Hamiltonian is equal to
\begin{eqnarray}
H_B=\frac{g_sl_s}{2}\tr \Pi_I\Pi_I-\frac{1}{4g_sl_s}\tr [\Phi^I,\Phi^J][\Phi^I,\Phi^J] \ . 
\end{eqnarray}
Then it is easy to see that  total  Hamiltonian that is given as linear combination of the bare Hamiltonian with the first class constraints  has the form
\begin{eqnarray}\label{totHam}
H_T=
\frac{g_sl_s}{2}\tr \Pi_I\Pi_I-\frac{1}{4g_sl_s}\tr [\Phi^I,\Phi^J][\Phi^I,\Phi^J]
+\lambda_I\bP_I+\lambda^{IJ}\bJ_{IJ} \  .
\nonumber \\
\end{eqnarray}
Since $\bP_I$ and $\bJ_{IJ}$ are first class constraints the standard procedure is to fix them. 
For example, we can impose the gauge fixing condition that says that the center of mass coordinates
are equal to zero. In other words we define gauge fixing functions $\mG_I$ as
\begin{equation}
	\mG_I\equiv \tr \Phi_I\approx 0 \ .
\end{equation}
Then we have
\begin{equation}
	\pb{\mG_I,\bP_J}=\pb{\Phi^I_{ij},(\Pi_J)_{kl}}\delta_{ji}\delta_{lk}=
\delta^I_J \delta_{ik}\delta_{jl}\delta_{ji}\delta_{lk}=\delta^I_J\delta_{il}\delta_{li}=
N\delta^I_J \ . 	
\end{equation}
In other words $\mG_I$ and $\bP_J$ are set of second class constraints that now strongly vanish.
We further fix generators $\bJ_{IJ}\approx 0$ by imposing conditions that off-diagonal components
of the matrix $I^{IJ}$ are zero 
\begin{equation}
	\mG^{IJ}\equiv I^{IJ}\approx 0 \ , I\neq J \ , 
\end{equation}
where
\begin{equation}
I^{IJ}=\frac{1}{2g_sl_s}\tr (\Phi^I\Phi^J)-\frac{1}{2g_s N l_s}\tr (\Phi^I)
\tr(\Phi^J)	 \ . 
\end{equation}
Note that we have following Poisson brackets
\begin{eqnarray}
\pb{\mG^{IJ},\bP_K}=0
\end{eqnarray}
together with 
\begin{eqnarray}
&&\pb{\mG^{IJ},\bJ^{KL}}=\frac{1}{2g_sl_s}(\delta^{IL}\Phi^J_{nm}\Phi_{mn}^K
-\delta^{IK}\Phi^{J}_{ji}\Phi^L_{ij}+\delta^{JL}\Phi^K_{mn}\Phi^I_{nm}-
\delta^{JK}\Phi^I_{mn}\Phi^L_{nm})-\nonumber \\
&&-\frac{1}{2g_sN}(\delta^{IL}\Phi^K_{ij}\delta_{ji}-\delta^{IK}\Phi_{ij}^L\delta_{ji})\tr\Phi^J-
\frac{1}{2g_sN}(\delta^{JL}\Phi^K_{ij}\delta_{ji}-\delta^{JK}\Phi_{ij}^L\delta_{ji})\tr\Phi^I=
\nonumber \\
&&=\delta^{IL}I^{JK}-\delta^{IK}I^{JL}+
\delta^{JL}I^{KI}-\delta^{JK}I^{IL} \ . \nonumber \\
\end{eqnarray}
For $I=L$ and $J=K$ we obtain non-zero result
\begin{equation}
	\pb{\mG^{LK},\bJ^{KL}}=I^{KK} \ . 
\end{equation}
Since $I^{KK}\neq 0$ by definition we find that   $\mG^{LK}$ is gauge fixing function for $\bJ^{KL}$ and they form collection of the second class constraints. 

It is important to stress that gauge fixed theory with $\bP_I=\bJ^{IJ}=0$ corresponds to the original Hamiltonian for $N$ D0-branes and we can interpret these frames with vanishing total momentum and angular momentum as Newtonian frames, for more details we recommend discussion presented in \cite{Ferraro:2014yza}.
\section{Gauging Time translation}\label{fourth}
In order to find an action invariant under arbitrary time dependent translation 
$t'=t+\epsilon(t)$ we follow the standard procedure of parametrized systems, see for example \cite{Henneaux:1992ig}. We begin with the canonical form of the action
\begin{equation}\label{orgact}
S=\int dt ((\Pi_I)_{ij}\dot{\Phi}_{ij}^I-H_T) \ ,
\end{equation}
where $H_T$ is given in (\ref{totHam}). 
As the next step we introduce variable $t$ and conjugate momenta $p_t$ and rewrite the action into the form
\begin{equation}\label{parsys}
S=\int d\tau (p_t\frac{d}{d\tau}t+
(\Pi_I)_{ij}\frac{d}{d\tau}\Phi_{ij}^I
-N(p_t+H_T)) \ . 
\end{equation}
In order to see equivalence between (\ref{parsys}) and (\ref{orgact}) let us consider equations of motion for $N$ and $p_t$ that give
\begin{equation}\label{sol}
\frac{d}{d\tau}t-N=0 \ , \quad  p_t+H_T=0 \ 
\end{equation}
that inserting back to the action (\ref{parsys}) we obtain
\begin{equation}\label{parsys1}
S=\int d\tau \frac{dt}{d\tau} ((\Pi_I)_{ij}\frac{d}{dt}\Phi_{ij}^I-H_T) \ , 
\end{equation}
where we presumed that the first relation in (\ref{sol}) can be inverted. Then it is easy to see that (\ref{parsys1}) is equivalent do (\ref{orgact}). The action (\ref{parsys}) is manifestly reparametrization invariant under transformation 
\begin{equation}
\tau=f(\tau') \ , \quad  t'(\tau')=t(\tau) \ , \quad  N(\tau)=N'(\tau')\frac{1}{\frac{df}{d\tau'}} \ 
\end{equation}
and 
\begin{equation}
(\Pi'_I)_{ij}(\tau')=(\Pi_I)_{ij}(\tau)  \ , \quad \Phi'^I_{ij}(\tau')=
\Phi_{ij}^I(\tau) \ . 
\end{equation}
In summary we got the action (\ref{parsys}) that is invariant under time dependent Galilean transformation together with arbitrary redefinition of the time $\tau$. Clearly this construction is generally correct even in our specific case of $N$ D0-branes. On the other hand the question is physical interpretation of the coordinate $t$ and 
how it should be interpreted in the context of non-abelian nature of D0-brane action. In other words this construction cannot be interpreted as covariant form of the action for $N$ D0-brane which is very difficult to construct, see for example \cite{Brecher:2004qi,Brecher:2005sj}.
 For that reason we mean that this construction has only formal meaning. 


 {\bf Acknowledgement:}

The work of J.K. was
supported by the Grant Agency of the Czech Republic under the grant
P201/12/G028.

	\end{document}